\documentstyle[aps]{revtex}
\begin{document}
\tightenlines
\title{Warm inflation with coupled thermal quantum fluctuations:
a new semiclassical approach\footnote{This essay received
an ``honorable mention'' in the
Annual Essay Competition of the Gravity Research Foundation
for the year 2000.}}
\author{Mauricio Bellini\footnote{E-mail: mbellini@mdp.edu.ar}}
\address{Departamento de F\'{\i}sica, Facultad de Ciencias Exactas
y Naturales, \\
Universidad Nacional de Mar del Plata, De\'an Funes, (7600)
Mar del Plata, Buenos Aires, Argentina}
\maketitle
\begin{abstract}
I consider a new semiclassical expansion for the inflaton field
in the framework of warm inflation scenario. The fluctuations
of the matter field are considered as thermally
coupled with the particles
of the thermal bath. This coupling parameter depends on the temperature
of the bath. 
The power spectrum remains invariant under this new semiclassical expansion
for the inflaton. I find that the thermal component of the amplitude for 
the primordial field fluctuations
should be very small at the end of inflation.
\end{abstract}
\vskip 2cm
\noindent
PACS number(s): 98.80.Cq.; 04.62.+v\\
\vskip 1cm
Inflation is needed because it solves the horizon, flatness, and monopole
problems of the very early universe. This theory also provides a mechanism
for the creation of primordial density fluctuations.
The differential microwave radiometer (DMR) on the Cosmic Background
Explorer (COBE) has made the first direct probe of the initial density
perturbations through detection
of the temperature anisotropies in the cosmic background radiation
(CBR). The results are consistent with the scaling spectrum
given by the inflationary model. For inflation the simplest
assumption is that there are two scales: a long - time, long - distance
scale associated with the vacuum energy dynamics and a single short - time,
short - distance scale associated with a random force component. The Hubble
time during inflation, $1/H$, appropriately separates the two regimes.

Quantum fluctuations\cite{1} and thermal fluctuations\cite{2}
of matter fields can play a prominent role in inflationary cosmology.
During inflation, vacuum fluctuations on scales smaller than the size
of the horizon are magnified into classical perturbations
on scales bigger than the Hubble radius\cite{3}.
The classical perturbations can lead to classical curvature of
spacetime and energy density
perturbations after inflation. These density perturbations should
be responsible for the formation of large - scale structure of the
universe, as well as, the anisotropies
in the cosmic microwave background\cite{4}.
Structure formation scenarios, can receive 
important restrictions based on the measured $\delta T_r/T_r = 1.1
\times 10^{-5}$. According to the standard inflationary model,
the formation of large - scale structure in the universe
has its origin in the growth of primordial inhomogeneities in the
matter distribution.

In a previous work\cite{5}
I studied a stochastic approach for
the backreaction of the metric produced by the fluctuations of the
matter field in the framework of warm inflation scenario.
In that work I considered a thermal coupling between
the scalar field and the fields in the thermal bath. The aim of this
letter is to consider a coupled
thermal interaction between the matter
field fluctuations and the fields in the thermal bath
to study the matter field fluctuations in
the warm inflation scenario.

Warm inflation
takes into account separately, the matter and radiation energy
fluctuations. In this scenario the field $\varphi$ interacts with the
particles of a thermal bath with a mean temperature
($T_r$) smaller
than the Grand Unified Theories (GUT) critical temperature:
$T_r \  < \  T_{GUT} \simeq 10^{15}$ GeV. This scenario was introduced by
A. Berera\cite{6,7}. In the
warm inflation era, the kinetic component of energy, $\rho_{kin}$, must
be smaller than the vacuum energy, which is given by the effective
potential $V(\varphi)$
\begin{equation}
\rho(\varphi) \sim \rho_m \sim V(\varphi) \gg \rho_{kin},
\end{equation}
where $\rho_{kin}(\varphi) = \rho_r(\varphi) + \dot\varphi^2/2$, and
the radiation energy density is $\rho_r(\varphi)
= {\tau(\varphi) \over 8 H(\varphi)} \dot\varphi^2$.
Here, $\varphi $ is a scalar field of matter,
$\tau(\varphi)$ is an effective friction parameter that represent
the interaction of the matter
field with other fields of the thermal bath\cite{BC}.

In this work I consider a new semiclassical expansion for
the inflaton field $\varphi$
\begin{equation}\label{se}
\varphi(\vec x,t) = \phi_c(t) + \alpha(t) \  \psi(\vec x,t),
\end{equation}
where $\phi_c(t) = <0|\varphi(\vec x,t)|0>$ and
$<0|\psi(\vec x,t)|0> = <0|\dot\psi(\vec x,t)|0> = 0$.
Here, $|0>$ is the vacuum state.
Here, $\alpha(t)$ is a dimensionless time - dependent function that
characterize the thermal coupling between the inflaton field
fluctuations
and the fields in the thermal bath. Furthermore,
the classical time - dependent field $\phi_c(t)$ gives the instantaneous
background. In principle, a permanent or temporary coupling
of the scalar field with other fields might also lead to dissipative
processis producing entropy at different eras of the cosmic evolution.
We introduce the function $\alpha(t)$ with the aim to
make an estimation of how
important is the thermal amplitude of the fluctuations
at the end of inflation.
I will consider $\alpha = \left({T_r(t) \over M}\right)^{\beta}$,
where $T_r(t)$ is the time - dependent radiation temperature and $M \simeq
10^{15}$ GeV is
the GUT mass. 

The dynamics of the classical field $\phi_c(t)$ was obtained in previous
works\cite{8}
\begin{equation}
\ddot\phi_c + \left[3 H_c+ \tau_c\right] \dot\phi_c + V'(\phi_c) =0,
\end{equation}
where $H_c \equiv H(\phi_c)=\dot a /a$, $\tau_c\equiv \tau(\phi_c)$ and
$V'(\phi_c) \equiv \left.{d V(\varphi) \over d \varphi}\right|_{\phi_c}$.
Furthermore $\dot\phi_c = - {M^2_p \over 4 \pi}
H'_c \left(1+ {\tau_c \over 3 H_c}\right)^{-1}$ and the
classical effective potential [see for example ref. (\cite{CQG})]
can be obtainded from the homogeneous Friedmann equation
$H^2_c(\phi_c) = {4\pi \over 3 M^2_p} \left[
\left(1+\tau_c/(4H_c)\right) \dot\phi^2_c + 2 V(\phi_c)\right]$:
\begin{equation}
V(\phi_c) = \frac{M^2_p}{8\pi} \left[ H^2_c -
\frac{M^2_p}{12\pi} (H'_c)^2
\left(1+\frac{\tau_c}{4 H_c}\right)\left(1+\frac{\tau_c}{3 H_c}\right)^{-2}
\right].
\end{equation}
The radiation energy density of the background is
\begin{equation}\label{r}
\rho_r[\phi_c(t)] \simeq \frac{\tau_c}{8 H_c}
\left(\frac{M^2_p}{4\pi}\right)^2
(H'_c)^2 \left(1+\frac{\tau_c}{3 H_c}\right)^{-2},
\end{equation}
and the temperature of the bath is $T_r \propto \rho^{1/4}_r[\phi_c(t)]$.
The spatially homogeneous matter energy density is given by
\begin{equation}
\rho_m(\phi_c) = \frac{\dot\phi^2_c}{2} + V(\phi_c),
\end{equation}
for $V(\phi_c) \gg  {\dot\phi^2_c \over 2}$.

I study the perturbations on a globally flat, homogeneous
and isotropic spacetime, described by a flat Friedmann - Robertson -
Walker (FRW)
metric $ds^2 = -dt^2 + a^2 d\vec x^2$.
The equation for the quantum perturbations $\psi$, with the semiclassical
expansion (\ref{se}), is
\begin{eqnarray}
\ddot\psi &+& \left[2\frac{\dot\alpha}{\alpha} + (3 H_c+\tau_c)\right]
\dot\psi - \frac{1}{a^2} \nabla^2\psi \nonumber \\
&+& \left[(3 H_c+\tau_c)
\frac{\dot\alpha}{\alpha} + \frac{\ddot\alpha}{\alpha}+
V''(\phi_c)\right] \psi = 0.
\end{eqnarray}
To simplify the structure of this equation, we can introduce the map
$\chi = e^{3/2 \int (H_c +\tau_c/3 + {2\dot\alpha \over 3 \alpha}) dt}
\psi$
\begin{equation}
\ddot\chi - \frac{1}{a^2} \nabla^2 \chi - \mu^2 \chi = 0,
\end{equation}
where $\mu^2(t) = {k^2_o(t) \over a^2}$ is a effective
squared time - dependent parameter of mass
\begin{equation}\label{ms}
\mu^2(t) = \frac{9}{4}\left
(H_c+\tau_c/3 \right)^2 - V''(\phi_c)
+ \frac{3}{2} \left(\dot H_c + \frac{\dot\tau_c}{3} \right). 
\end{equation}
Note that this parameter does not depends on $\alpha$.
For $H_c +\tau_c/3
+ {2\dot\alpha \over 3 \alpha}=0$, one obtains the particular map
$\psi_{cg} = \chi_{cg}$. Furthermore, for $\tau_c/3+2 \dot\alpha/(3\alpha)=0$ the
resulting map is $\psi_{cg} = e^{-3/2 \int H_c dt} \  \chi_{cg}$,
which coincides with the map developed in standard inflation\cite{3}
with the standard
semiclassical expansion $\varphi = \phi_c + \phi$.

The backreaction of the metric with the quantum fluctuations of the
matter field
introduces an effective curvature ($K$) on the
globally flat FRW background metric, which appears in the semiclassical
Friedmann equation
\begin{equation}
H^2_c + \frac{K}{a^2} = \frac{8 \pi}{3 M^2_p}
\left<E\left|\rho_m(\varphi) + \rho_r(\varphi)\right|E\right>,
\end{equation}
where
\begin{eqnarray}
\frac{K}{a^2} &= & \frac{8 \pi}{3 M^2_p} \left[ \left(1+
\frac{\tau_c}{8 H_c}\right) \left(\frac{\dot\alpha^2}{2} <\psi^2> +
\frac{\alpha^2}{2} <\dot\psi^2> \right.\right. \nonumber \\
& + & \left.\left. \alpha \dot\alpha <\psi \dot\psi>\right)
+ \frac{\alpha^2}{a^2} \left<\left(\vec \nabla \psi\right)^2 \right>
+ \frac{V''}{2} \alpha^2 \left<\psi^2\right> \right]. \label{cur}
\end{eqnarray}

To study the consequences of this approach, we can
consider a power - law expansion of the universe.
In this model the scale factor is $a \propto (t/t_o)^p$, and
$H_c(t) = p/t$.
The effective classical potential and the radiation energy
density are given by
\begin{eqnarray}
V[\phi_c(t)] &=& \frac{3 M^2_p}{8\pi} t^{-2} \nonumber \\ &\times
& \left[p^2 - \frac{M^2_p}{2\pi m^2} p^2 \left(1+\frac{\tau_c t}{4
p}\right) \left(1+ \frac{\tau_c t}{3 p} \right)^{-2}\right], \\
\rho_r[\phi_c(t)] & = & \left(\frac{p \  \tau_c \  t}{32}\right)
\left(\frac{M^2_p}{\pi m}\right)^2 \left(1+\frac{\tau_c t}{3
p}\right)^{-2} t^{-2},
\end{eqnarray}
where $p/t = H_o e^{\phi_c(t)/m}$.

The redefined matter field perturbations on
the infrared sector [$k^2 \ll k^2_o(t)$] takes into account
only the modes much bigger than the size of the horizon. It can
be written as a Fourier expansion in terms of the modes
$\chi_k(\vec x,t)=e^{i \vec k . \vec x} \xi_k(t)$
\begin{equation}\label{chi}
\chi_{cg} = \frac{1}{(2\pi)^{3/2}} \int d^3 k \  \theta(\epsilon k_o - k)
\left[ a_k \chi_k + a^{\dagger}_k \chi^*_k\right],
\end{equation}
where $a_k$ and $a^{\dagger}_k$ are the annihilation and creation
operators with commutation relations $[a_k,a^{\dagger}_{k'}] =
\delta^{(3)}(\vec{k}-\vec{ k'})$
and $[a^{\dagger}_k,a^{\dagger}_{k'}]$ $=$
$[a_k,a_{k'}] = 0$. The dimensionless constant
$\epsilon = k/k_o \ll 1$ is introduced to take
into account only the modes
with wavelengths much bigger than the size of the horizon.
As was showed in a previous work\cite{CQG}, the redefined fluctuations
$\chi_{cg}$ are classical in the infrared sector. This implies
that $\xi_k \dot\xi^*_k - \dot\xi_k \xi^*_k \simeq 0$.

To make a calculation in the power - law inflation model I
consider $\alpha = [T_r(t)/M]^{\beta}$ (with $\beta \ge 0$
and $\alpha <1$),
and $\tau_c = \gamma (p/t)$.
Here, $M\simeq 10^{15}$ GeV,
is the GUT mass. 
However, the form chosen for the 
function $\alpha$ must come from the analysis
of some quantum field theoretical 
description of the coupling between the scalar
field and the radiation fields which would result in the introduction
of an affective temperature dependent parameter in the expansion
of the field.
The effective squared
parameter of mass (\ref{ms}), for $\rho_r = {\pi^2 \over
30} N(T_r) \  T^4_r$ --- where $N(T_r)$ is the 
number of relativistic degrees of freedom at
temperature $T_r$ --- is given by
\begin{eqnarray}
\mu^2(t) &=& \frac{t^{-2}}{12}\left[ 3p^2(9+6\gamma+\gamma^2)- 6 p
(3+\gamma)+36\gamma+144\right. \nonumber \\ 
&-& \left.
\frac{m^2}{M^2_p} \left(288 \pi +192
\pi\gamma+32\pi\gamma^2\right) \right].\label{mu}
\end{eqnarray}
The effective potential and the homogeneous component of the
radiation energy density are, in this case
\begin{eqnarray}
V(\phi_c) & = &
\frac{3M^2_p}{8\pi} H^2_o e^{2\phi_c/m}
\left[1- \frac{M^2_p}{2\pi m^2} \left(1+\frac{\gamma}{4}\right)
\left(1+\frac{\gamma}{3}\right)^{-2}\right], \\
\rho_r(\phi_c) & = & \frac{\gamma}{32} \left(1+\frac{\gamma}{3}\right)^{-2}
\left(\frac{M^2_p}{\pi m}\right)^2 H^2_o \  e^{2\phi_c/m},
\end{eqnarray}
which increase exponentially on $\phi_c$. Note that
$\dot\rho_r <0$, which means that $\dot\alpha <0$.

The equation of
motion for the time - dependent modes $\xi_k$ is
\begin{equation}\label{xi}
\ddot\xi_k+ \left[\frac{k^2}{a^2} - \mu^2(t)\right] \xi_k = 0,
\end{equation}
where the squared time - dependent parameter of mass is given by
eq. (\ref{mu}). 
As $\mu(t)$ is independent of $\alpha$ [see eq. (\ref{ms})], there are no
changes in the dynamics of the functions $\xi_k$.
The asymptotic solution for eq. (\ref{xi})
(for large $p$-values), is\cite{NP}
\begin{equation}
\xi_k(t) \simeq \frac{\sqrt{t/t_o}}{2\sqrt{2\pi}}
\Gamma(\nu) \left[ \frac{k(t/t_o)^{1-p}}{2(p-1)H_o}\right]^{-\nu},
\end{equation}
where $\Gamma(\nu)$ is the gamma function,
$\nu = {1\over 2(p-1)} \sqrt{1+ 4 L^2}$ and
\begin{eqnarray}
L^2 &=& \frac{1}{12} \left[3 p^2 (9+6\gamma +\gamma^2) - 6 p (
3+\gamma) \right. \nonumber \\
&+& \left.144  - \frac{m^2}{M^2_p} (288 \pi
+ 192 \pi\gamma + 32 \pi\gamma^2)\right].
\end{eqnarray}
The squared fluctuations for $\psi_{cg}$ $=$
$e^{-3/2 \int (H_c +\tau_c/3
+ {2\dot\alpha \over 3 \alpha}) dt} \  \chi_{cg}$ are given by
\begin{equation}
\left< \psi^2_{cg}(\vec x,t)\right>_{IR} =
e^{-3 \int (H_c +\tau_c/3 + {2\dot\alpha \over 3 \alpha}) dt}
\int^{k_o}_{0} \frac{d k \  k^2}{(2 \pi^2)} \xi^2_k(t).
\end{equation}
As the coupling parameter
$\alpha(t)$ decreases during the rapid expansion of the universe
[$\dot\alpha(t) <0$], the sign of the effective curvature $K$
will be dependent on
the term $\alpha \dot\alpha <\phi \dot\phi>$ in eq. (\ref{cur}).
However, during inflation the increasing rate of the
scale factor is bigger than the $K$ one, so that at the end of
inflation the ratio ${K \over a^2}$ becomes nearly zero.

From the condition $n-1 = 2(1-\nu)$ for the spectral index
$n$\cite{NP}, one
obtains the following
condition for the constraint $|n-1| < 0.3$ obtained with the
COBE data
\begin{equation}\label{L}
\frac{\left[0.702 (p-1)\right]^2 - 1}{4}
< L^2  <  \frac{\left[1.3 (p-1)\right]^2 - 1}{4}.
\end{equation}
A spectral index $n \sim 1$, also
has been found in a model
with cosmic strings plus cold or hot dark matter\cite{AS,be3}.
The standard choice of $n=1$ was first advocated by Harrison\cite{ha} and
Zel'dovich\cite{Ze} on the ground that it is scale invariant
at the epoch of horizon entry.
The condition (\ref{L}) can be written as
\begin{equation}
\gamma_1(p) < \gamma < \gamma_2(p),
\end{equation}
where $\gamma_1(p)$ and $\gamma_2(p)$ are given by the expressions
\begin{eqnarray}
&&\gamma_1(p) =\frac{1}{f_1(p)} \left[10^4 \left( 75 p M^2_p -
225 p^2 M^2_p + 2400 m^2 \pi\right)\right. \nonumber \\
&& + 500 M_p \left( 23654592 m^2 \pi - 106891191 p^2 M^2_p 
\right. \nonumber \\
&&  - 2217618 p^3 M^2_p 
+1108809 p^4 M^2_p
-11827296 p^2 m^2 \pi \nonumber \\
&& \left.\left. + 1164172704 m^2 \pi\right)^{1/2}\right] \\
&& \gamma_2(p) = \frac{1}{f_2(p)} \left[ 300 \left( [M^2_p - 3 M^2_p +32
m^2\pi\right) \right. \nonumber \\
&& + 10 M_p \left( 32448 p m^2 \pi
-41679 p^2 M^2_p - 3042 p^3 M^2_p \right.\nonumber \\
&& \left.\left. + 1521 p^4 M^2_p 
- 16224 p^2 m^2 \pi +
454176 m^2 \pi\right)^{1/2} \right],
\end{eqnarray}
with
\begin{eqnarray}
f_1(p) & = & 750000 p^2 M^2_p - 8000000 m^2 \pi, \\ 
f_2(p) & = & 300 p^2 M^2_p - 3200 m^2 \pi.
\end{eqnarray}

Finally, we can calculate the dependence in the amplitude of the
fluctuations due to the semiclassical expansion here introduced.
If $\left<\psi^2_{cg}\right>^{(\beta \neq 0)}_{IR}$ is
the amplitude for the fluctuations
of the matter field within thermal coupling 
(i.e., with $\beta \neq 0$) and 
$\left<\psi^2_{cg}\right>^{(\beta= 0)}_{IR}$ is the amplitude for the
matter field fluctuations without thermal coupling 
(i.e., for $\beta = 0$), the only
contribution in the amplitude for the fluctuations
that arises from the  thermal coupling
will be
\begin{equation}\label{rat}
R(t) = \frac{\left<\psi^2_{cg}\right>^{(\beta= 0)}_{IR}}{
\left<\psi^2_{cg}\right>^{(\beta \neq 0)}_{IR}} =
\left(\frac{T_r(t)}{M}\right)^{2\beta},
\end{equation}
where
\begin{eqnarray}
\left(\frac{T_r(t)}{M}\right)^2 &=& 
\frac{M^2_p}{8\pi M^2} \sqrt{\frac{15 \gamma}{N}}
\left(1+\frac{\gamma}{3}\right)^{-1}\nonumber \\
& = & \frac{M^2_p}{8\pi M^2} \sqrt{\frac{15 \gamma}{N}} \frac{p}{m}
\left(1+\frac{\gamma}{3}\right)^{-1} \  t^{-1} ,
\end{eqnarray}
due to $T_r(t) \sim \rho^{1/4}_r(t)$.
The eq. (\ref{rat}) defines the ratio
between the thermal and matter field fluctuations. 
For $\beta >  0$, one obtains
that this ratio decreases with time as $t^{-\beta}$, and 
thus the thermal fluctuations should be more small
than the matter fluctuations at the end of inflation.
Furthermore, as $\left({T_r(t)\over M}\right)^2 \ll 1$, one
obtains that $\tau_c/H_c < 10^{-15}$.
Here, $\beta $ should be calculared from quantum field theory.
(see for example ref. \cite{st})).

To summarize, in this work I considered a new semiclassical
expansion for the inflaton field in warm inflation. The matter
field fluctuations are considered as thermally
coupled with the particles
of the thermal bath with temperature $T_r < T_{GUT}$. The coupling
parameter $\alpha = (T_r/M)^{\beta}$, depends 
on the temperature of the bath.
In this framework, I find that the equation of motion for the
redefined matter field fluctuations $\chi$, remains invariant
under the transformation $\chi = e^{3/2 \int(H_c+\tau_c/3+ {2\dot\alpha
\over 3 \alpha})dt} \psi$. This implies that the chosen semiclassical
expansion $\varphi = \phi_c + \alpha(t) \psi$ does not modifies
the power spectrum --- obtained with the standard semiclassical
expansion $\varphi = \phi_c + \phi$ --- for the fluctuations 
of energy density. The thermal contribution to the
fluctuations are given by the ratio $R = \alpha^2$.
I find --- studying a power - law expansion for the universe ---
that the thermal component of the amplitude for these
fluctuations in the infrared sector, should be very small at
the end of inflation.


\begin{thebibliography}{99}
\bibitem{1} A. A. Starobinsky, in {\em Fundamental Interactions}, ed.
V. N. Ponomarev (MGPI Press, Moscow, 1984). P. 54.\\
A. A. Starobinsky, in {\em Current Topics in Field Theory
Quantum Gravity, and Strings}, ed. by H. J. de Vega and
N. S\'anchez, Lecture in Physics, p. 107 (Springer, New York,
1986).
\bibitem{2} Arjun Berera and Li-Zhi Fang, Phys. Rev. Lett. {\bf 74},
1912 (1995).
\bibitem{3} M. Bellini, H. Casini, R. Montemayor and P. Sisterna,
Phys. Rev. {\bf D54}, 7172 (1996).
\bibitem{4} V. F. Mukhanov, L. R. W. Abramo and R. H. Branderberger,
Phys. Rev. Lett. {\bf 78}, 1624 (1997).
\bibitem{5} M. Bellini, Class. Quantum Grav. {\bf 17}, 145 (2000). 
Preprint gr-qc/9910073. 
\bibitem{6} A. Berera, Phys. Rev. Lett. {\bf 18}, 3218 (1995).
\bibitem{7} A. Berera, Phys. Rev. {\bf D54}, 2519 (1996);
A. Berera, M. Gleiser and R. O. Ramos, Phys. Rev. Lett. {\bf 83},
264 (1999).
W. Lee and Li-Zhi Fang, Phys. Rev. {\bf D59}, 083503 (1999).
\bibitem{BC} the reader can find an overview about Interactions
in scalar field cosmology in: A. P. Billyard and A. A. Coley,
Phys. Rev. {\bf D61}, 083503 (2000).
\bibitem{8} M. Bellini, Phys. Lett. {\bf B428}, 31 (1998);
M. Bellini, Nuovo Cim. {\bf B113}, 1481 (1998).
\bibitem{CQG} M. Bellini, Class. Quant. Grav. {\bf 16}, 2393 (1999).
\bibitem{NP} M. Bellini, Nucl. Phys. {\bf B563}, 245 (1999).
\bibitem{AS} A. Albrech and A. Stebbins, Phys. Rev. Lett. {\bf 68}, 2121;
{\bf 69}, 2615 (1992).
\bibitem{be3}
A. Berera, L. Z. Fang, and G. Hinshaw, Phys. Rev. {\bf D57}, 2207 (1998).
\bibitem{ha} R. Harrison, Phys. Rev. {\bf D}, 2726 (1970).
\bibitem{Ze} Ya. B. Zel'dovich, Astron. Astrophys. {\bf 5}, 84 (1970).
\bibitem{st}
A. Berera, M. Gleiser and R. O. Ramos, Phys. Rev. {\bf D58}, 123508
(1998).

\end{thebibliography}
\end{document}